\documentclass[twocolumn,preprintnumbers,floats,prd,amssymb,floatfix,nofootinbib,balancelastpage,superscriptaddress,amsmath]{revtex4-1}

\pdfoutput=1
\usepackage[utf8]{inputenc}
\usepackage{CJKutf8}
\usepackage{amssymb}
\usepackage{amsmath}
\usepackage{amsfonts}
\usepackage{graphicx}
\usepackage{color}
\usepackage{xspace}
\usepackage{comment}
\usepackage{hyperref}
\usepackage[normalem]{ulem}
\usepackage[section]{placeins}
\usepackage{afterpage}
\usepackage{slashed}
\usepackage{enumerate}
\usepackage{enumitem}
\usepackage{caption}
\usepackage{subfigure}
\usepackage{float}
\usepackage{ulem}
\usepackage{verbatim}
\usepackage{multirow,rotating}
\usepackage[dvipsnames]{xcolor}
\usepackage{braket}
\usepackage{booktabs} 
\usepackage{array}     

\definecolor{dcolour}{rgb}{.5, .5, .5}
\def\gsim{\raise0.3ex\hbox{$\;>$\kern-0.75em\raise-1.1ex\hbox{$\sim\;$}}}
\def\lsim{\raise0.3ex\hbox{$\;<$\kern-0.75em\raise-1.1ex\hbox{$\sim\;$}}}
\def\gsim{\raise0.3ex\hbox{$\;>$\kern-0.75em\raise-1.1ex\hbox{$\sim\;$}}}
\def\lsim{\raise0.3ex\hbox{$\;<$\kern-0.75em\raise-1.1ex\hbox{$\sim\;$}}}

\newcommand{\ba}[1]{\begin{eqnarray} \label{(#1)}}
	\newcommand{\ea}{\end{eqnarray}}

\newcommand{\Eq}[1]{Eq.~\ref{#1}}
\newcommand{\FIG}[1]{FIG.~\ref{#1}}
\newcommand{\TAB}[1]{Table~\ref{#1}}
\begin{document}
\captionsetup[figure]{justification=raggedright,singlelinecheck=false}
\title{Theoretical study of the $N(1535)$ in the process $\Lambda_c^+\to n\pi^+\pi^0$}
\author{Ruitian Li}
\email{liruitian@mail.dlut.edu.cn}
\affiliation{Institute of Theoretical Physics, School of Physics, Dalian University of Technology, \\ 
		No.2 Linggong Road, Dalian, Liaoning, 116024, People’s Republic of China}
\author{Xuan Luo}
\email{xuanluo@ahu.edu.cn}
\affiliation{School of Physics and Optoelectronics Engineering, Anhui University, \\
		Hefei, Anhui 230601, People’s Republic of China}
\author{Hao Sun}
\email{haosun@dlut.edu.cn}
\affiliation{Institute of Theoretical Physics, School of Physics, Dalian University of Technology, \\ 
	No.2 Linggong Road, Dalian, Liaoning, 116024, People’s Republic of China}
\begin{abstract}
	We have investigated the $\Lambda_c^+\to n\pi^+\pi^0$ decay process using the chiral unitary approach, by considering that the nucleon resonance $N(1535)$ could be dynamically generated through $S$-wave pseudoscalar meson-octet baryon interactions. In the invariant mass distributions of $\pi^0n$ and $\pi^+n$, we observe a distinct peak structure associated with the resonance states $N(1535)^0$ and $N(1535)^+$, respectively. Our results suggest that the $\Lambda_c^+\to n\pi^+\pi^0$ process can be utilized to probe the properties of the nucleon $N(1535)$. Therefore, we encourage more precise experimental measurements of this process in the future.
\end{abstract}
\vskip10 mm
\maketitle
\flushbottom	

\section{Introduction}
\label{I}

The low-lying excited baryons have aroused much attention since their being discovered, but the problem of their internal structure still needs more theoretical and high-statistic experimental studies to be determined. In particular, understanding the nature of the low-excited baryons with quantum number $J^P=1/2^-$, is one of the most challenging topics. For example, in the standard quark model~\cite{Capstick:2000qj}, $N(1535)$ as the first orbital excitation of the nucleon with negative parity should have lower energy than the first radial excitation $N(1440)$ with positive parity, but the experimental result is contrary~\cite{ParticleDataGroup:2024cfk}, which is also known as the ``mass inversion puzzle''. This implies that $N(1535)$ may have a complex internal structure and there has been a lot of studies~\cite{Glozman:1997ag,Bijker:1994yr,Helminen:2000jb,An:2008xk,Bijker:2015gyk} that have attempted to solve the problem with the description of the nature of $N(1535)$ by some extensions to the conventional quark model.

Many researches~\cite{Liu:2005pm,Geng:2008cv,Xie:2017erh,Pavao:2018wdf,ParticleDataGroup:2022pth} have shown that the coupling strength of $N(1535)$ to $\eta N$ and $K\Lambda$ is significant, and also been shown that $N(1535)$ is couples strongly to $\eta' N$ by studying the $\gamma p \to p\eta'$~\cite{CLAS:2005rxe} and $pp\to pp\eta'$~\cite{Cao:2008st} reactions, and large coupling of $N(1535)$ to $\phi N$ by studying the $\pi^- p \to n\phi$, $pp\to pp\phi$ and $pn\to d\phi$ reactions~\cite{Xie:2007qt,Doring:2008sv,Cao:2009ea}, which indicate that there is a large $s\bar{s}$ component in the $N(1535)$ resonance state.  

Thus, Ref.~\cite{Zou:2007mk,Zhang:2004xt,Hannelius:2000gu} have suggested that $N(1535)$ can be interpreted as a mixture of some pentaquark configuration and three quark components of $uud$, which would natural explanation the heavier mass of  $N(1535)$ than $N(1440)$ and the large couplings to the strangeness channel.

Many studies also suggest that the structure of $N(1535)$ can be interpreted as a molecular picture. In Ref.~\cite{Kaiser:1995eg,Kaiser:1996js}, the chiral unitary approach was first employed to describe $N(1535)$ as the bound state of $K\Lambda$ and $K\Sigma$. Other studies~\cite{Nieves:2001wt,Bruns:2019fwi,Nacher:1999vg} have also supported that $N(1535)$ as a molecular picture in the coupled channels. Recently, it has also been studied to explain the molecular nature of $N(1535)$ by measuring the correlation functions~\cite{Molina:2023jov}, or the scattering length and effective range of the $K\Sigma$, $K\Lambda$ and $\eta p$ channels~\cite{Li:2023pjx}. In Ref.~\cite{Abell:2023nex,Guo:2022hud,Liu:2015ktc}, it was proposed, using Hamiltonian Effective Field Theory, that $N(1535)$ could be a three quark core dressed by meson-baryon scattering contributions.

Moreover, $N(1535)$ could also be dynamically generated via $S$-wave pseudoscalar meson-octet baryon interactions within the chiral unitary approach, and is predicted strongly coupled to the $\pi N$, $\eta N$, $K\Lambda$ and $K\Sigma$ channels. In the $J^P=1/2^-$ sector, the physical picture stays unchanged when considering pseudoscalar meson-baryon mixing with vector meson-baryon states~\cite{Garzon:2014ida}and is in a good agreement with most experimental values. In addition, there are many researches that have obtained the masse and width of $N(1535)$ from the position of the pole on the second Riemann sheet using the above method that are in agreement with the experimental results~\cite{Inoue:2001ip,Bruns:2010sv,Nieves:2011gb,Gamermann:2011mq,Khemchandani:2013nma,Garzon:2014ida}.

Recently, numerous studies on the multi-body nonleptonic weak decays of charmed baryons have proven to be an exceptional platform for investigating hadron resonances~\cite{Miyahara:2015cja,Wang:2022nac,Xie:2017erh,Hyodo:2011js,Wang:2020pem,Zeng:2020och,Feng:2020jvp,Oset:2016lyh}, as these processes exhibit a large phase space and involve complex final state interactions. Until now, the nature of the $N(1535)$ resonance remains undetermined,  so it is necessary to continue the exploration of it~\cite{Crede:2013kia,Klempt:2007cp}. Thus, in this work, we study the charm baryon of $\Lambda_c^+\to n\pi^+\pi^0$ decay process based on the chiral unitary approach, which can dynamically generate the $N(1535)$ resonance by considering the $S$-wave pseudoscalar meson-octet baryon interactions. Then, we obtain the invariant mass distributions of $\pi^0n$ and $\pi^+n$, and the appearance of peaks linked to $N(1535)$ is evident from the results.

This paper is organized as follows. First, in Sec.~\ref{II}, we present the theoretical formalism of the process $\Lambda_c^+\to n\pi^+\pi^0$. Then, in Sec.~\ref{III}, we discuss numerical results and discussions. Finally, we provide a brief summary in Sec.~\ref{IV}.
\section{formalism}
\label{II}

\begin{figure}[tbhp]
	\centering
	\includegraphics[scale=0.3]{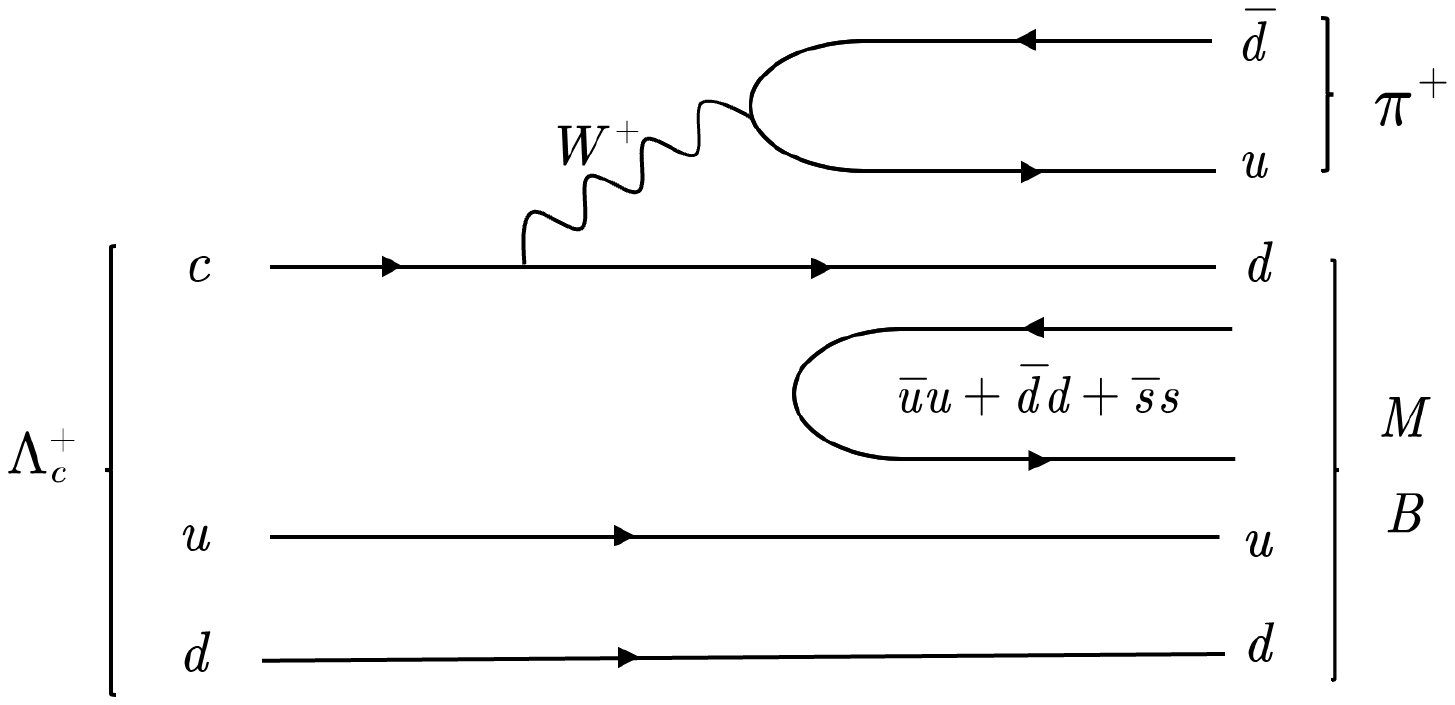}
	\caption{Quark level diagram for the process $\Lambda_c^+ \to \pi^+ d \left(\bar{u}u+\bar{d}d+\bar{s}s\right)ud$ via the $W^+$ external emission.}
	\label{fig1}
\end{figure}

\begin{figure}[tbhp]
	\centering
	\includegraphics[scale=0.3]{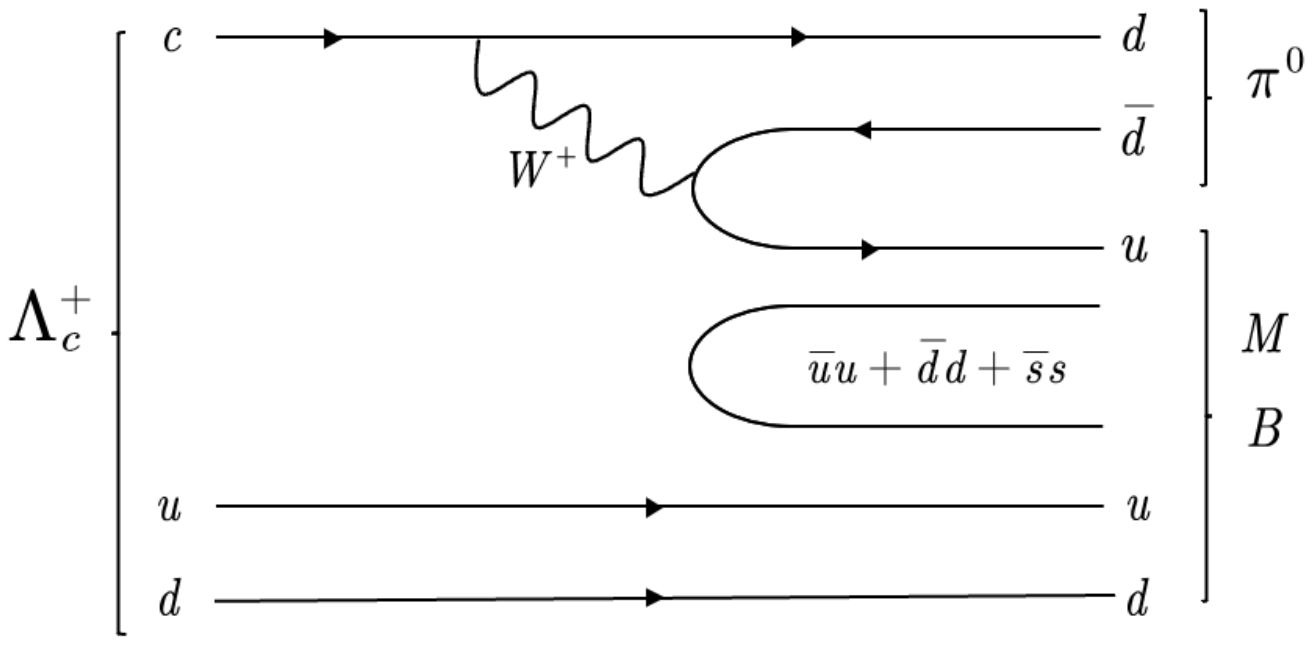}
	\caption{Quark level diagram for the process $\Lambda_c^+ \to \pi^0 u \left(\bar{u}u+\bar{d}d+\bar{s}s\right)ud$ via the $W^+$ internal emission.}
	\label{fig2}
\end{figure}

In this section, we present the theoretical formalism for the process $\Lambda_c^+\to n\pi^+\pi^0$.
We have considered $N(1535)$ produced by the dynamics of the meson-baryon interaction, as in \cite{Xie:2017erh,Li:2024rqb,Lyu:2023aqn,Li:2025exm,Li:2025gvo}.
In this work, for the $\Lambda_c^+ \to \pi^+(MB)\to\pi^+\pi^0n$ process, we consider the $W^+$ external emission as shown in \FIG{fig1}, and for the $\Lambda_c^+ \to \pi^0(MB)\to\pi^0 \pi^+n$ process, we consider the $W^+$ internal emission, as illustrated in \FIG{fig2}.
For \FIG{fig1}, the initial state $\Lambda_c^+$ of the $c$ quark first weakly decays into a $d$ quark and a $W^+$ boson. Subsequently, $W^+$ boson weakly decays to the $u\bar{d}$ quark pair.  The $u\bar{d}$ quark pair is then hadronized to $\pi^+$, while the remaining quark cluster $dud$ and the quark pair $\bar{q}q=\bar{u}u+\bar{d}d+\bar{s}s$ produced in the vacuum with quantum numbers $J^{PC}=0^{++}$ hadronize into meson-baryon(MB) pairs. And for \FIG{fig2}, after the initial state $\Lambda_c^+$ of $c$ quark weakly decays to a $d$ quark and a $W^+$ boson, the $W^+$ boson then weakly decays to $u\bar{d}$ quark pair. The $d\bar{d}$ quark pair is then hadronized to $\pi^0$, while the remaining $uud$ quark cluster and a $\bar{q}q$ quark pair produced in the vacuum are then hadronized to meson-baryon pairs. These two processes can be expressed as follows:

\begin{equation}
	\begin{aligned}
		\Lambda_c^+(1)&\Rightarrow \frac{1}{\sqrt{2}}c\left(ud-du\right)\\
		&\Rightarrow \frac{1}{\sqrt{2}}W^+ d\left(ud-du\right)\\
		& \Rightarrow \frac{1}{\sqrt{2}}u\bar{d}d\left(u\bar{u}+d\bar{d}+s\bar{s}\right)\left(ud-du\right)\\
		& \Rightarrow \frac{1}{\sqrt{2}}\pi^+ d\left(u\bar{u}+d\bar{d}+s\bar{s}\right)\left(ud-du\right),
	\end{aligned}
\end{equation}
\begin{equation}
	\begin{aligned}
		\Lambda_c^+(2)&\Rightarrow \frac{1}{\sqrt{2}}c\left(ud-du\right)\\
		&\Rightarrow \frac{1}{\sqrt{2}}W^+ d\left(ud-du\right)\\
		& \Rightarrow \frac{1}{\sqrt{2}}d\bar{d}u\left(u\bar{u}+d\bar{d}+s\bar{s}\right)\left(ud-du\right)\\
		& \Rightarrow \frac{1}{\sqrt{2}}\pi^0 u\left(u\bar{u}+d\bar{d}+s\bar{s}\right)\left(ud-du\right),
	\end{aligned}
\end{equation}

where we use the baryon flavor wave function $\Lambda_c^+ = \frac{1}{\sqrt{2}}c\left(ud-du\right)$. Based on the $SU(3)$ flavor symmetry, we connect the two degrees of freedom of quarks and hadrons using the meson matrix $M$ and the baryon matrix $B$. The specific forms of the $M$ and $B$ matrices~\cite{Bramon:1992kr,Molina:2019udw,Lyu:2023aqn} are as follows:
\begin{eqnarray}
	M&=&\left(\begin{matrix} u\bar{u} & u\bar{d} & u\bar{s}  \\
		d\bar{u}  &   d\bar{d}  &  d\bar{s} \\
		s\bar{u}  &  s\bar{d}   &    s\bar{s}
	\end{matrix}
	\right)   \\  
	&=&\left(\begin{matrix} \frac{\eta}{\sqrt{3}}+ \frac{{\pi}^0}{\sqrt{2}}+ \frac{{\eta}'}{\sqrt{6}} & \pi^+ & K^+  \\
		\pi^-  &   \frac{\eta}{\sqrt{3}}- \frac{{\pi}^0}{\sqrt{2}}+ \frac{{\eta}'}{\sqrt{6}}  &  K^0 \\
		K^-  &  \bar{K}^{0}   &    -\frac{\eta}{\sqrt{3}}+ \frac{{\sqrt{6}\eta}'}{3}
	\end{matrix}
	\right),\nonumber
\end{eqnarray}
\begin{eqnarray}
	B &=& \frac{1}{\sqrt{2}}\begin{pmatrix} 
		u(ds-sd) & u(su-us) & u(ud-du)  \\
		d(ds-sd) & d(su-us) & d(ud-du) \\
		s(ds-sd) & s(su-us) & s(ud-du)
	\end{pmatrix} \\
	&=& \begin{pmatrix} 
		\frac{\Sigma^0}{\sqrt{2}} + \frac{{\Lambda}}{\sqrt{6}} & \Sigma^+ & p  \\
		\Sigma^- & -\frac{\Sigma^0}{\sqrt{2}} + \frac{{\Lambda}}{\sqrt{6}}  & n \\
		\Xi^-  & \Xi^{0} & -\frac{2\Lambda}{\sqrt{6}} 		\end{pmatrix},
	\nonumber
\end{eqnarray}
where we have employed the $\eta-\eta'$ mixing as in Ref.~\cite{Bramon:1994cb,Lyu:2024qgc}. The hadronization processes can be expressed using the meson matrix $M$ and baryon matrix $B$:
\begin{equation}
	\begin{aligned}
		\Lambda_c^+ (1)& \Rightarrow \frac{1}{\sqrt{2}}\pi^+d\left(u\bar{u}+d\bar{d}+s\bar{s}\right)\left(ud-du\right)\\
		& = \pi^+ \sum M_{2i}B_{i3}\\
		& = \pi^+\left(|\pi^- p\rangle +\frac{1}{\sqrt{3}}|\eta n\rangle -\frac{1}{\sqrt{2}}|\pi^0 n\rangle -\frac{2}{\sqrt{6}}|K^0\Lambda\rangle\right),
	\end{aligned}
\label{eq5}
\end{equation}
\begin{equation}
	\begin{aligned}
		\Lambda_c^+ (2)& \Rightarrow \frac{1}{\sqrt{2}}\pi^0u\left(u\bar{u}+d\bar{d}+s\bar{s}\right)\left(ud-du\right)\\
		& = \pi^0 \sum M_{1i}B_{i3}\\
		& = \pi^0\left( \frac{1}{\sqrt{3}}|\eta p\rangle+ \frac{1}{\sqrt{2}}|\pi^0 p \rangle+ |\pi^+ n\rangle -\frac{2}{\sqrt{6}}|K^+\Lambda\rangle \right),
	\end{aligned}
\label{eq6}
\end{equation}
and since the large mass of $\eta'$, the $\eta' n$ and $\eta' p$ channels are not taken into account.
Considering the isospin multiplets $N=(p,n)$ and $\pi=(-\pi^+,\pi^0,\pi^-)$, we can then express the physical states in \Eq{eq5} and \Eq{eq6} as isospin states,
\begin{figure}[tbhp]
	\centering
	\includegraphics[scale=0.27]{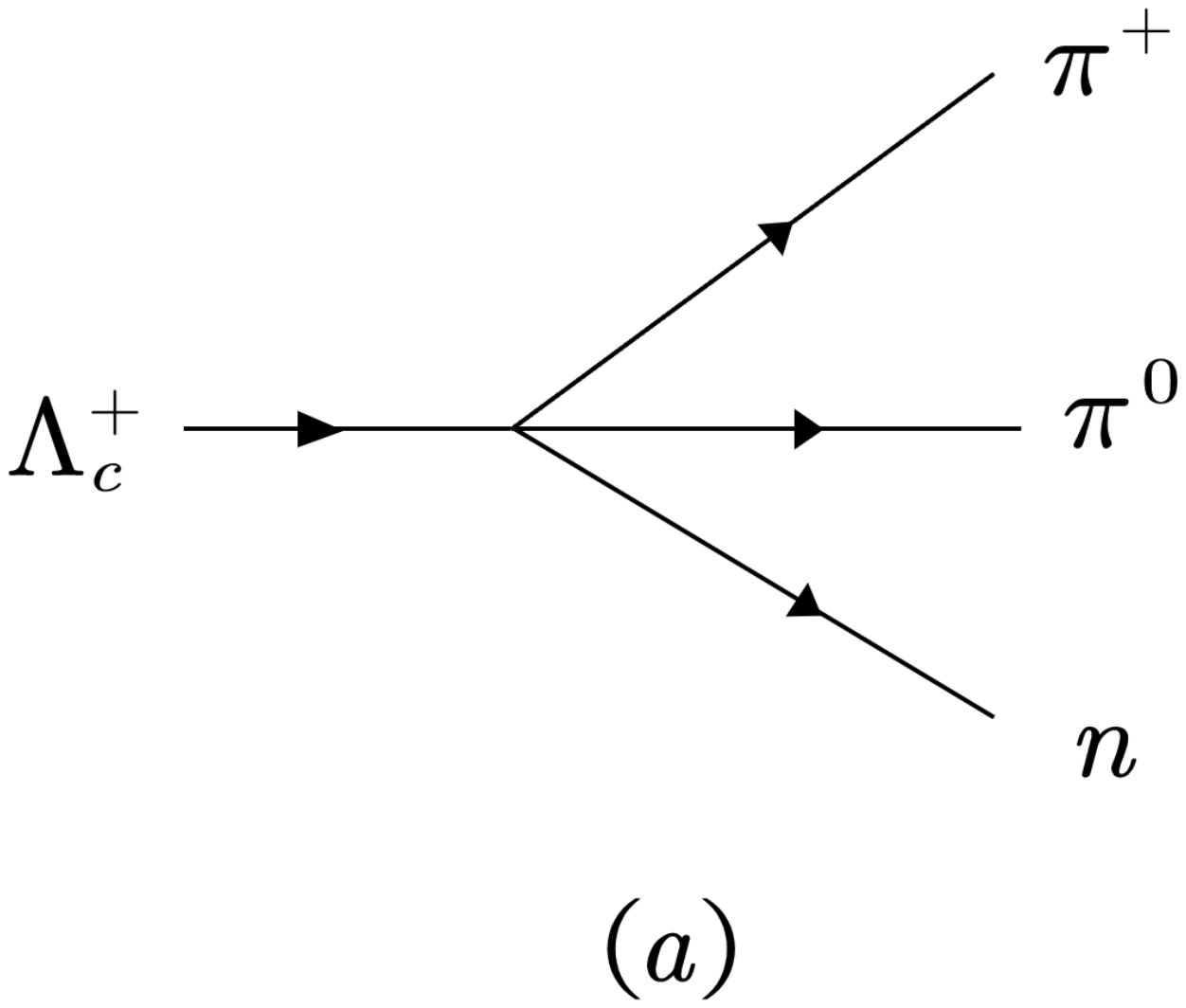}\\
	\includegraphics[scale=0.27]{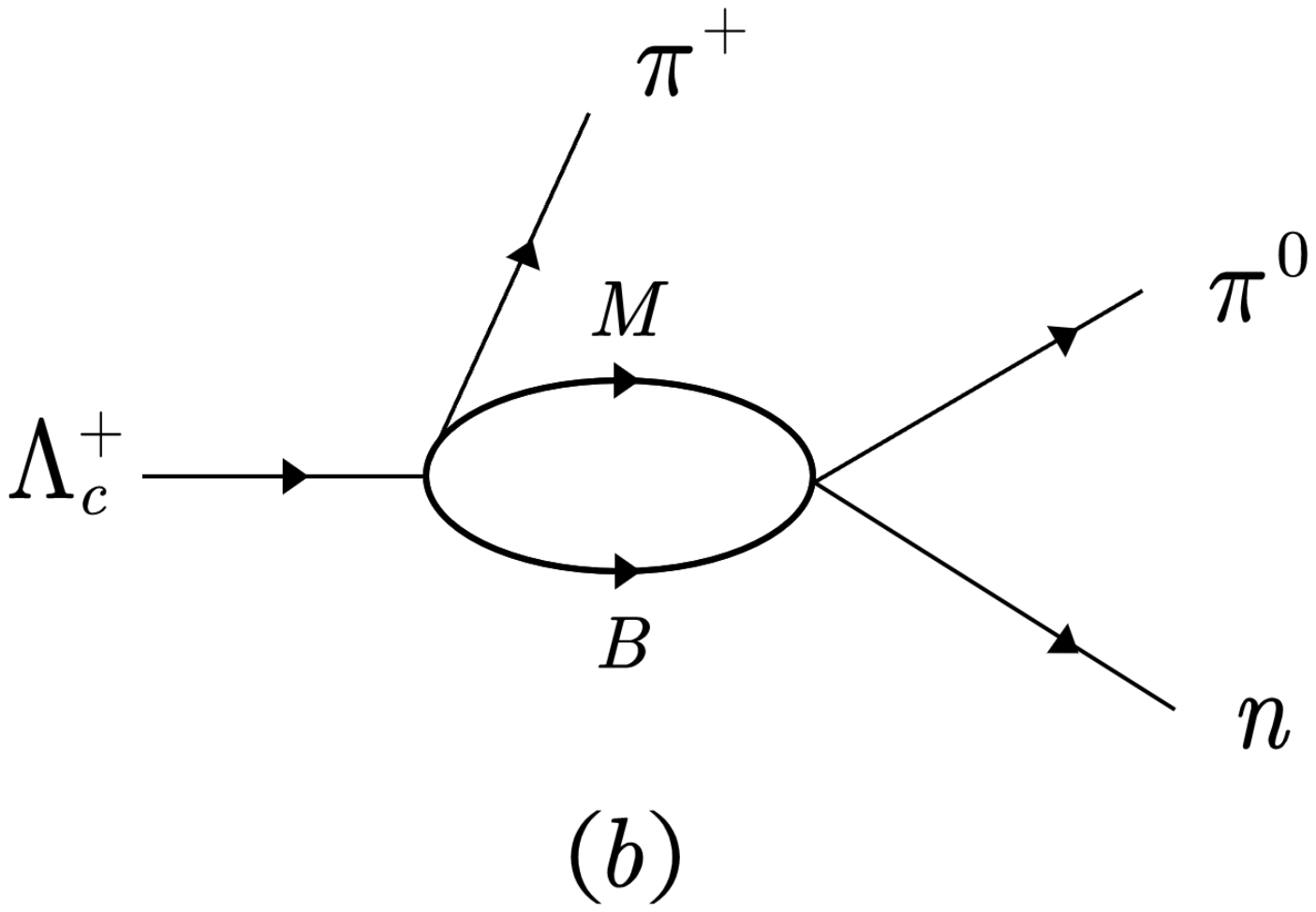}
	\caption{Diagrams for the process $\Lambda_c^+ \to \pi^+\pi^0 n$. (a) Tree level diagram, (b) final$\pi^0n$ rescattering.}
	\label{fig3}
\end{figure}
\begin{figure}[tbhp]
	\centering
	\includegraphics[scale=0.25]{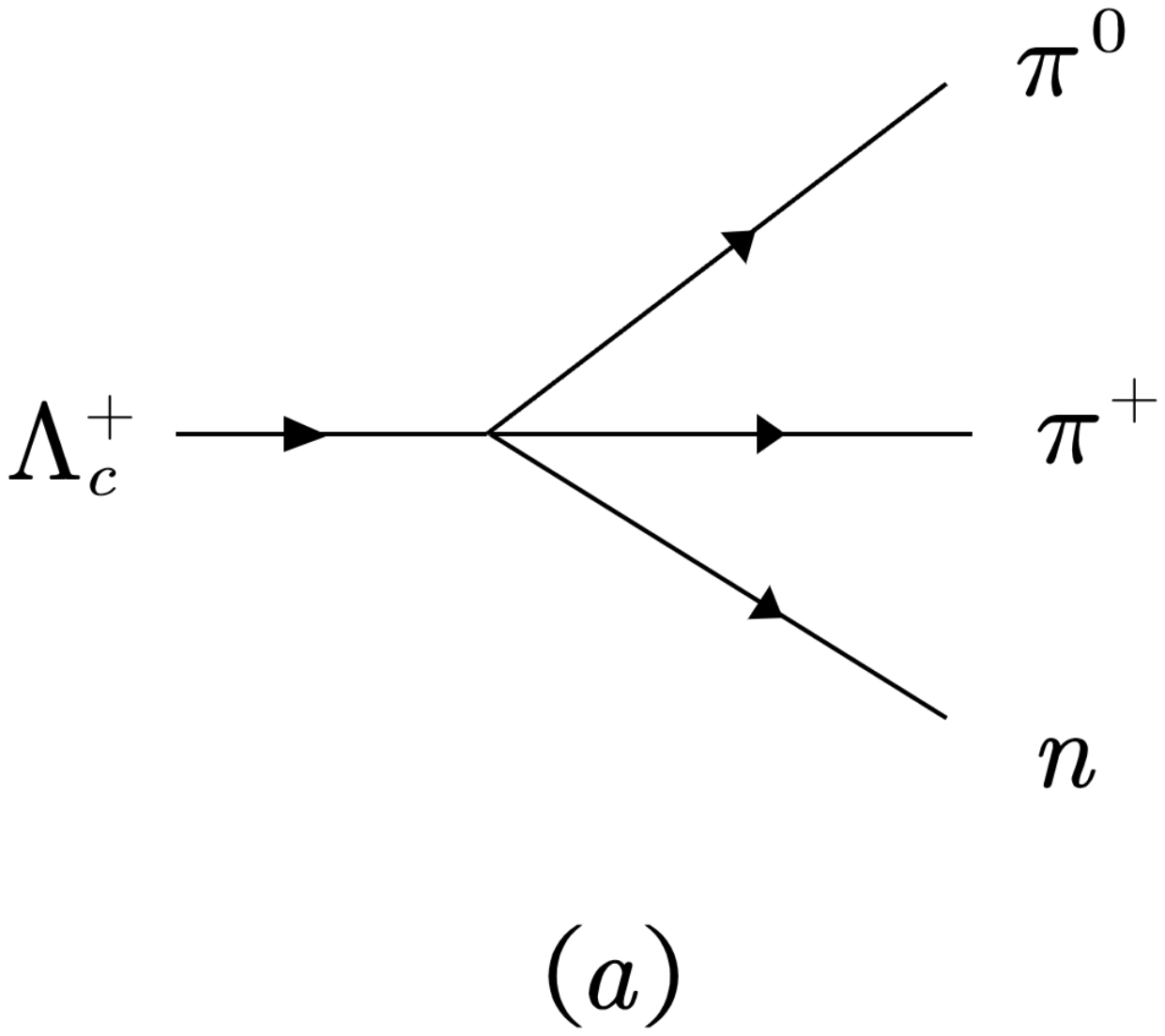}\\
	\includegraphics[scale=0.25]{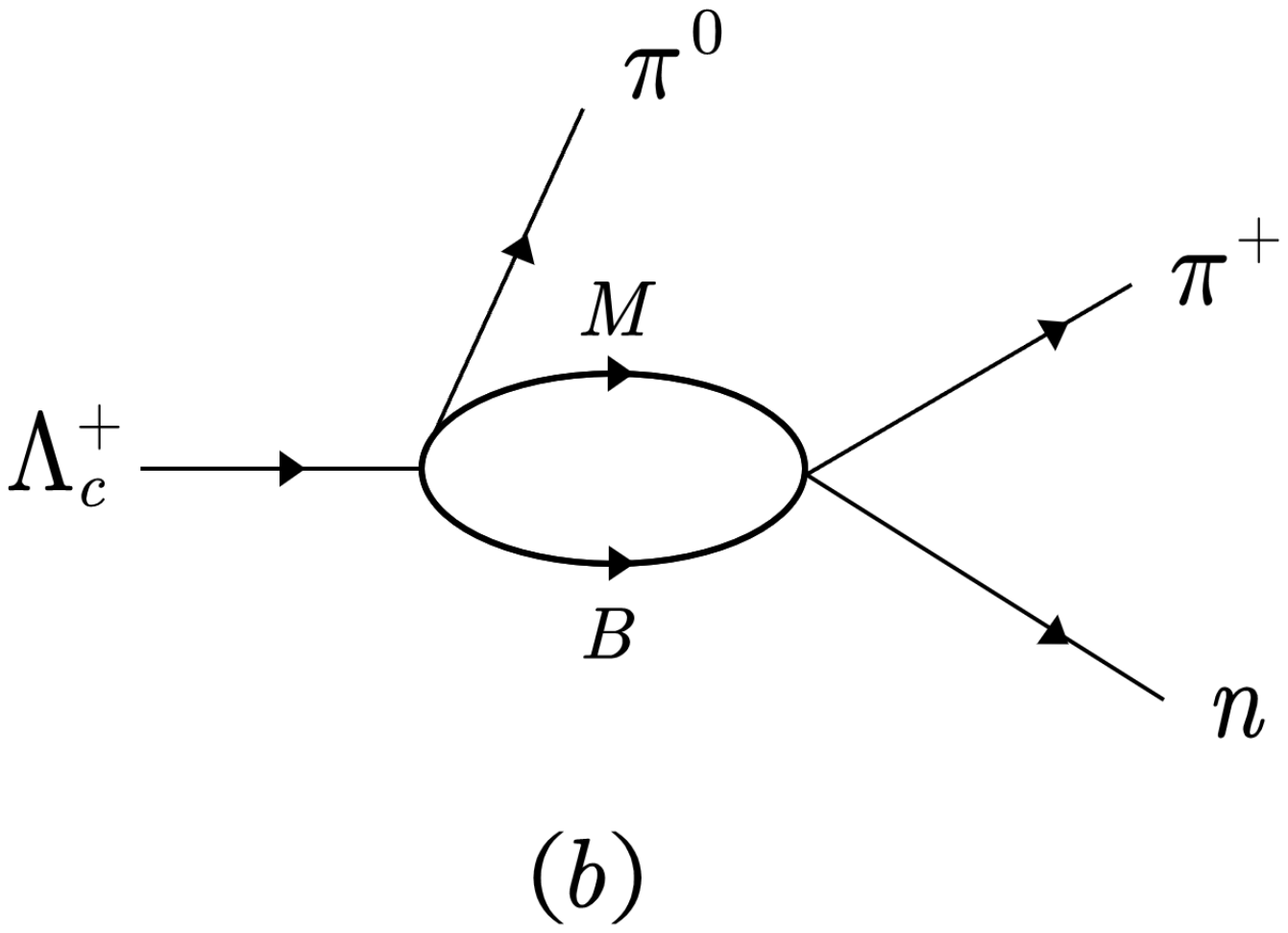}
	\caption{Diagrams for the process $\Lambda_c^+ \to \pi^+\pi^0 n$. (a) Tree level diagram, (b) $\pi^+n$ rescattering.}
	\label{fig4}
\end{figure}

\begin{equation}
	\begin{aligned}
		|\pi^- p \rangle & = |1,\frac{1}{2};-1,\frac{1}{2}\rangle = -\sqrt{\frac{2}{3}}|\pi N\rangle^{I=\frac{1}{2}} +\sqrt{\frac{1}{3}}|\pi N\rangle^{I=\frac{3}{2}}\\
		|\pi^0 n\rangle& = |1,\frac{1}{2};0-\frac{1}{2}\rangle=\sqrt{\frac{1}{3}}|\pi N\rangle^{I=\frac{1}{2}}+\sqrt{\frac{2}{3}}|\pi N\rangle^{I=\frac{3}{2}} \\
		|\pi^0 p\rangle & = |1,\frac{1}{2}; 0,\frac{1}{2} \rangle =-\sqrt{\frac{1}{3}}|\pi N\rangle^{I=\frac{1}{2}}+\sqrt{\frac{2}{3}}|\pi N\rangle^{I=\frac{3}{2}}\\
		|\pi^+ n\rangle& = -|1,\frac{1}{2};1,-\frac{1}{2}\rangle =-\sqrt{\frac{2}{3}}|\pi N\rangle^{I=\frac{1}{2}}-\sqrt{\frac{1}{3}}|\pi N\rangle^{I=\frac{3}{2}}.
	\end{aligned}
\end{equation}
As $N(1535)$ with the isospin quantum number of $I=\frac{1}{2}$, we only consider contributions from the $I=\frac{1}{2}$ region. In the isospin basis, \Eq{eq5} and \Eq{eq6} are written as:
\begin{equation}
	\begin{aligned}
		\Lambda_c^+(1)&
	 =  \pi^+\left(-\sqrt{\frac{3}{2}}|\pi N\rangle +\frac{1}{\sqrt{3}}|\eta N\rangle -\frac{2}{\sqrt{6}}|K\Lambda\rangle\right)\\
	  \Lambda_c^+ (2)& = \pi^0 \left(-\sqrt{\frac{3}{2}}|\pi N\rangle +\frac{1}{\sqrt{3}}|\eta N\rangle -\frac{2}{\sqrt{6}}|K\Lambda\rangle\right).
	\end{aligned}
\end{equation}

The meson-baryon final state interactions dynamically generating the $N(1535)$ resonance, and the amplitude of the process $\Lambda_c^+ \to \pi^+ \pi^0 n$ for the contributions of \FIG{fig3} and \FIG{fig4} is:

\begin{equation}
	\begin{aligned}
		\mathcal{T}^{MB}(1) =&CV_p\left[
		-\sqrt{\frac{3}{2}}-\sqrt{\frac{3}{2}}G_{\pi N}(M_{\pi^0n})t_{\pi N\to \pi N}(M_{\pi^0n}) \right. \\
		&\left. +\frac{1}{\sqrt{3}} G_{\eta N}(M_{\pi^0n})t_{\eta N \to \pi N}(M_{\pi^0n})\right. \\
		&\left.-\frac{2}{\sqrt{6}} G_{K\Lambda}(M_{\pi^0n})t_{K\Lambda\to \pi N}(M_{\pi^0n})\right],\\
		\mathcal{T}^{MB}(2) =&V_p\left[
		-\sqrt{\frac{3}{2}}-\sqrt{\frac{3}{2}}G_{\pi N}(M_{\pi^+n})t_{\pi N\to \pi N}(M_{\pi^+n}) \right. \\
		&\left. +\frac{1}{\sqrt{3}} G_{\eta N}(M_{\pi^+n})t_{\eta N \to \pi N}(M_{\pi^+n})\right. \\
		&\left.-\frac{2}{\sqrt{6}} G_{K\Lambda}(M_{\pi^+n})t_{K\Lambda\to \pi N}(M_{\pi^+n})\right],
	\end{aligned}\label{eq8}
\end{equation}
where the color factor $C$ represents the relative weight of the external $W^+$ emission mechanism in \FIG{fig1} compared to internal $W^+$ emission in \FIG{fig2}, the parameter $V_p$ represents the strength of the weak interaction vertex in \FIG{fig3} and \FIG{fig4}, which we treat as a constant in the present study.

In \Eq{eq8}, $G_l$ stands for the meson-baryon loop function~\cite{Inoue:2001ip}
\begin{equation}
	G_l(s) = i\int \frac{d^4q}{(2\pi)^4}\frac{2M_l}{(P-q)^2-M_l^2+i\epsilon}\frac{1}{q^2-m_l^2+i\epsilon},
	\label{eq9}
\end{equation}
where $M_l$ and $m_l$ denote the masses of the baryon and meson in the $l$-th channel, respectively, and $P$ and $q$ are the four-momentum of this meson-baryon system and the meson in the center of mass system, respectively. As for the loop function $G_l$ in \Eq{eq9}, the integral exhibits logarithmic divergence. Typically, we may adopt either the three-momentum cutoff method or dimensional regularization scheme~\cite{Inoue:2001ip,Guo:2005wp}. In this work, we employ the three-momentum cutoff method to obtain the following analytic expression for loop function $G_l$,
\begin{equation}
	\begin{aligned}
		G(s) = & \frac{2M_l}{16\pi^2 s} \Bigg\lbrace \sigma \left( \arctan \frac{s + \Delta}{\sigma \lambda_1} + \arctan \frac{s - \Delta}{\sigma \lambda_2} \right) \\
		& - \left[ (s + \Delta) \ln \frac{(1 + \lambda_1) q_{\text{max}}}{M_l} \right.\\
		&\left.+ (s - \Delta) \ln \frac{(1 + \lambda_2) q_{\text{max}}}{m_l} \right] \Bigg\rbrace ,
	\end{aligned}
\end{equation}
with
\begin{align}
	\sigma&=\sqrt{[-(s-(M_l+m_l)^2)(s-(M_l-m_l)^2)]},\\
	\Delta&=M_l^2-m_l^2,\\
	\lambda_1&=\sqrt{1+\frac{M_l^2}{q^2_{\text{max}}}}, \qquad \lambda_2=\sqrt{1+\frac{m_l^2}{q^2_{\text{max}}}}.
\end{align}
To dynamically generate $N(1535)$ resonance, we take the cut-off parameter $ q_{\text{max}}= 1150\text{MeV}$, consistent with Ref.~\cite{Li:2024rqb,Li:2025gvo}.  For $\sqrt{s} > m_1+m_2$, the pole of the possible bound state is searched in the second Riemann sheet, where we employ $G^{II}$~\cite{Li:2024rqb}:

\begin{equation}
	G^{II}(\sqrt{s})=G(\sqrt{s})+i\frac{|\vec{q}|}{4\pi \sqrt{s}}, \qquad \text{Im}(|\vec{q}|)>0,
	\label{eq15}
\end{equation}
where $\vec{q}$ is the momentum of the meson in the centre of mass frame.
Since the $\pi N$-channel threshold is far away from the considered energy region of $N(1535)$ that we are mainly focused on, in this case, as done in Ref.~\cite{Dai:2020yfu,Oset:2024fbk}, we take only the imaginary part of the loop function, which has a simple analytic form within the second part in \Eq{eq15}.

The two-body scattering amplitude $t_{MB-\pi N}$ of the coupled channel in \Eq{eq8} is calculated by solving the Bethe-Salpeter equation through the method of chiral unitary approach:
\begin{equation}
	T=[1-VG]^{-1}V.
\end{equation}
We have included four coupled channels in our calculation: $\pi N$, $\eta N$, $K\Lambda$ and $K\Sigma$.
The $V$ is transition potential, which is expressed as in Ref.~\cite{Oset:2001cn,Wang:2015pcn},
\begin{equation}
	\begin{aligned}
		V_{ij}=&-C_{ij}\frac{1}{4f_if_j}(2\sqrt{s}-M_i-M_j)\sqrt{\frac{E_i+M_i}{2M_i}}\sqrt{\frac{E_j+M_j}{2M_j}},
	\end{aligned} 
\end{equation}
where, $M_{i(j)}$ and $E_{i(j)}$ represent the mass and energy of the baryon in the $i(j)$-th channel, respectively, and $E_i = (s+M_i^2-m_i^2)/2\sqrt{s}$. The coefficient $C_{ij}$, which reflect $SU(3)$ flavor symmetry, is listed in \TAB{tab1}. The decay constant $f_{i(j)}$ for the meson in the $i(j)$-th channel is:
\begin{equation}
	\begin{aligned}
			f_{\pi}=93\ \text{MeV}, \qquad f_{K} = 1.22f_{\pi}, \qquad f_{\eta} = 1.3f_{\pi} .
	\end{aligned}
\end{equation}

\begin{table}[htbp]
	\centering
	\caption{The $S\text{-wave}$ meson-baryon scattering coefficients \cite{Oset:1997it,Li:2024rqb}.}
	\setlength{\tabcolsep}{12pt}     
	\begin{tabular}{ccccc}
		\toprule 
		& $\pi N$ & $\eta N$ & $K\Lambda$ & $K\Sigma$ \\
		\midrule 
		$\pi N$  & 2        & 0        & $3/2$     & $-1/2$    \\
		$\eta N$ &          & 0        & $-3/2$    & $-3/2$    \\
		$K\Lambda$ &        &          & 0         & 0         \\
		$K\Sigma$ &         &          &           & 2         \\
		\bottomrule
	\end{tabular}
	\label{tab1}
\end{table}
{\tiny }

\begin{figure}[tbhp]
	\centering
	\includegraphics[scale=0.5]{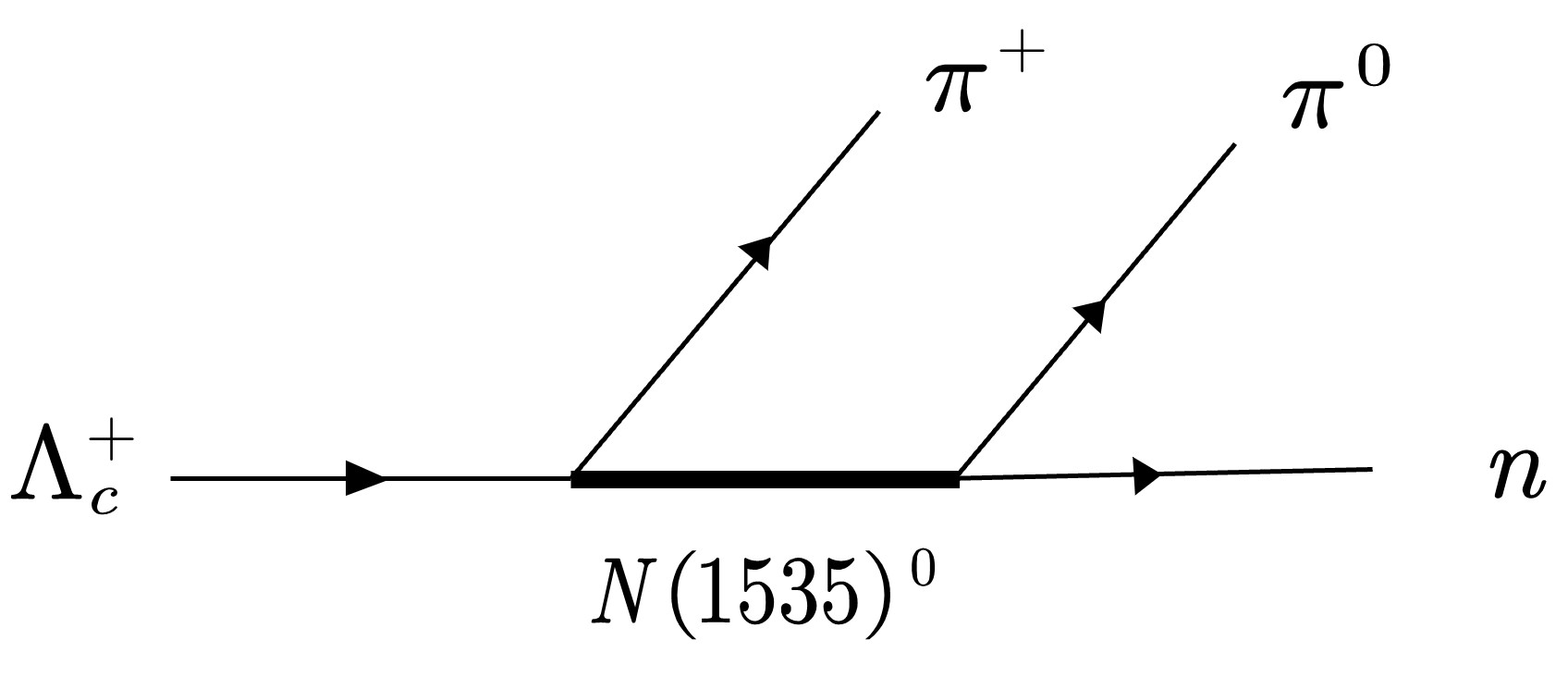}
	\caption{Diagram representing the process $\Lambda_c^+ \to$ $\pi^+N(1535)^0\to\pi^+\pi^0 n$.}
	\label{BW}
\end{figure}
Furthermore, in order to differentiate between the molecular and genuine baryon scenarios, we have also investigated the Breit-Wigner amplitude of the $N(1535)^0$ resonance contribution using the effective Lagrangian approach. The Breit-Wigner decay amplitude is shown in \FIG{BW}. 
The decay amplitude of $\Lambda_c^+\to \pi^+N(1535)^0$ consists of two kinds of structures~\cite{Xie:2017erh}:
\begin{equation}
	\mathcal{M}=i\bar{u}(q)(A+B\gamma_5)u(p)
\end{equation}
where $p$, $p_1$, $q$, $p_2$, and $p_3$ denotes the momentum of $\Lambda_c^+, \pi^+, N(1535)^0, \pi^0$ and $n$, respectively. 
Due to the lack of understanding of the properties of the $N(1535)$ resonance, we are currently unable to determine the coefficients $A$ and $B$ for the charmed baryons into meson and baryon states, and therefore we adopt $A = B$ in this study. We use the effective Lagrangian commonly used in Ref.~\cite{Zou:2002yy,Lu:2014yba} for the $\pi N N(1535)$ vertex:
\begin{equation}
	\mathcal{L}_{\pi N}^{N(1535)} = ig_{\pi NN^*} \bar{N}\vec{\tau} \cdot \vec{\pi} N^*+ h.c.
\end{equation}
where $N$, $N^*$ represent the fileds of $n$ and $N(1535)$ resonance, respectively.

The invariant decay amplitude of the $\Lambda_c^+ \to \pi^+N(1535)^0\to \pi^+\pi^0 n$ decay is:
\begin{equation}
	\begin{aligned}
			\mathcal{\tilde{T}}=&ig_{\pi NN^*}\bar{u}(p_3)\frac{\slashed{q}+M_{N(1535)}}{q^2-M_{N(1535)}^2+iM_{N(1535)}\Gamma_{N(1535)}(q^2)}\\
			&(A+B\gamma_5)u(p)
	\end{aligned}
\end{equation}
where $M_{N(1535)}$ and $\Gamma_{N(1535)}(q^2)$ are the mass and total decay width of $N(1535)$, respectively.
For the $N(1535)$ resonance, we implement the momentum dependent width $\Gamma_{N(1535)}$.
Since $\pi N$ and $\eta N$ are the main decay channels of $N(1535)$, we also use the following form as used in Refs.~\cite{Xie:2017erh,Wu:2009nw,Xie:2013wfa}:
\begin{equation}
	\Gamma_{N(1535)}(q^2)= \Gamma_{N^*\to \pi N}(q^2)+\Gamma_{N^*\to \eta N}(q^2) +\Gamma_0
\end{equation}
where we label $N(1535)$ as $N^*$ and $\Gamma_{N^*\to \pi N}(q^2)$ and 	$\Gamma_{N^*\to \eta N}(q^2)$ are the following forms, respectively,
\begin{equation}
	\begin{aligned}
		\Gamma_{N^*\to \pi N}(q^2)&=\frac{3g^2_{N^*N\pi}}{4\pi}\frac{\sqrt{|\vec{p}_{N\pi}|+m_n^2}+m_n}{\sqrt{q^2}}|\vec{p}_{N\pi}|,\\
		\Gamma_{N^*\to \eta N}(q^2)&=\frac{g^2_{N^*N\eta}}{4\pi}\frac{\sqrt{|\vec{p}_{N\eta}|+m_n^2}+m_n}{\sqrt{q^2}}|\vec{p}_{N\eta}|.
	\end{aligned}
\end{equation}
In this work, we take $g^2_{N^*N\pi}/ 4\pi=0.037$, $g^2_{N^*N\eta}/ 4\pi=0.28$ and $\Gamma_0=19.5$ MeV as used in Ref.~\cite{Xie:2017erh}.

With the theoretical formalism presented above, the total decay amplitude for the process of $\Lambda_c^+\to n\pi^+\pi^0$ can be expressed as,
\begin{equation}
	|\mathcal{T}|^2 = |\mathcal{T}^{MB}(1)+\mathcal{T}^{MB}(2)|^2.
	\label{eq19}
\end{equation}
We can calculate the double differential width according to the following equation~\cite{ParticleDataGroup:2022pth}:
\begin{eqnarray}
	\frac{d^2\Gamma}{dM^2_{\pi^+n}{dM^2_{\pi^0n}}}&=&\frac{4M_{\Lambda_c^+}M_n}{{(2\pi)}^3{32M_{\Lambda_c^+}}^3}|\mathcal{T}|^2. \label {eq:dgammadm12dm23} \nonumber \\
\end{eqnarray}
For a given value of $M_{12}$, the range of $M_{23}$ is constrained by the following condition:
\begin{eqnarray}
	M^{\rm max}_{23} &= &\sqrt{\left(E_{2}^\ast+E_{3}^\ast\right)^2 -\left(\sqrt{E_{2}^{\ast2}-m_{2}^2}-\sqrt{E_{3}^{\ast2}-m_{3}^2}\right)^2}, \nonumber \\
	M_{23}^{\rm min} &=&\sqrt{\left(E_{2}^\ast+E_{3}^\ast\right)^2 -\left(\sqrt{E_{2}^{\ast2}-m_{2}^2}+\sqrt{E_{3}^{\ast2}-m_{3}^2}\right)^2},\nonumber \\
\end{eqnarray}
where $E_{2}^\ast$ and $E_{3}^\ast$ are the energies of particles 2 and 3 in the $M_{12}$ rest frame, which are written as
\begin{align}
	&E_{2}^\ast=\frac{M_{12}^2-m_{1}^2+m_{2}^2}{2M_{12}},  \nonumber \\
	&E_{3}^\ast=\frac{M_{\Xi_c^+}^2-M_{12}^2-m_{3}^2}{2M_{12}},
\end{align}
with $m_1$, $m_2$ and $m_3$ are the masses of involved particles $1, 2$, and 3, respectively.

Here we need to note that the effective range of the chiral unitary approach is limited for calculating the differential mass distributions.  In order to make our predictions more reliable, following Ref.~\cite{Debastiani:2016ayp,Duan:2024czu}, and similarly we multiply the $Gt$ term of scattering amplitude of  in the high energy region by a smooth factor that decreases as the invariant mass $M_{\text{inv}}$ increases:
\begin{equation}
	Gt(M_{\text{inv}}) = Gt(M_{\text{cut}})e^{-\alpha(M_{\text{inv}}-M_{\text{cut}})}, \quad \text{for}\ M_{\text{inv}} > M_{\text{cut}}.
	\label{eq23}
\end{equation}
In this work, we adopt  $M_{\text{cut}} = 1650$ MeV and the value of $\alpha$ we will discuss in Sec.~\ref{III}.

\section{results and discussions}
\label{III}
\begin{figure}[tbhp]
	\centering
	\includegraphics[scale=0.45]{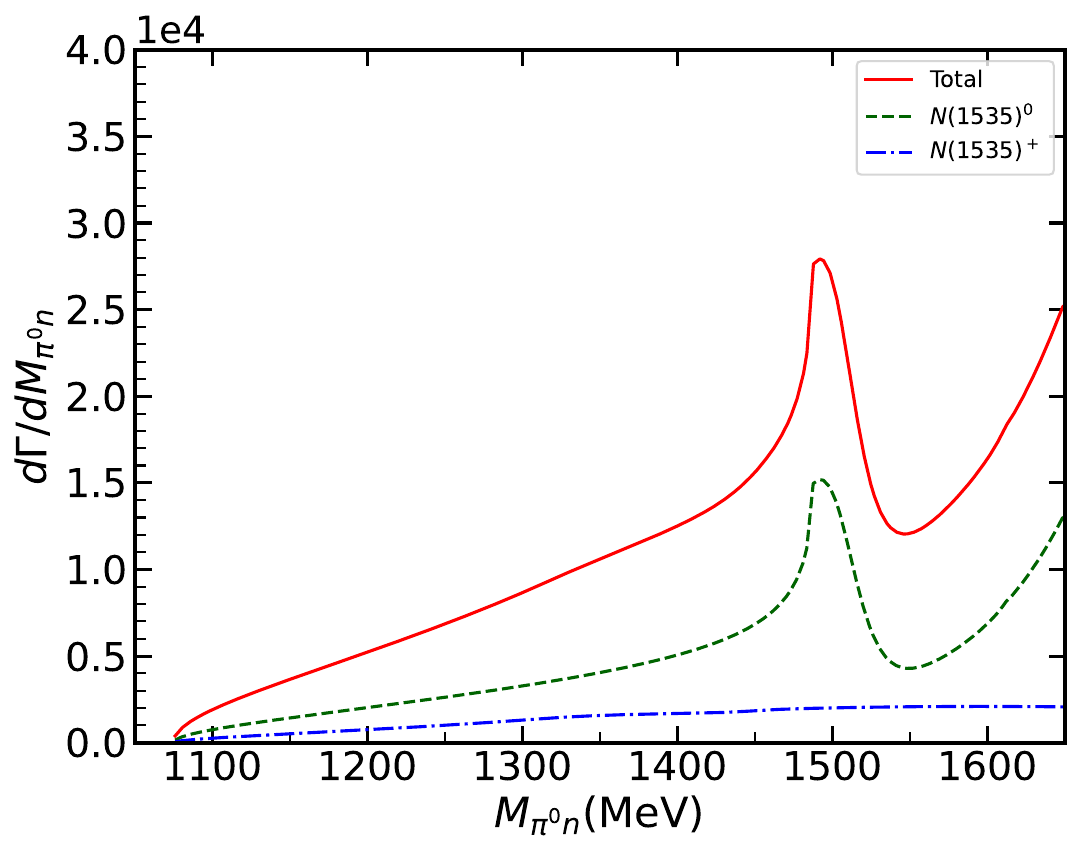}
	\caption{The $\pi^0 n$ invariant mass distribution of the process $\Lambda_c^+ \to \pi^+\pi^0 n$ decay.}
	\label{min23}	
\end{figure}
\begin{figure}[tbhp]
	\centering
	\includegraphics[scale=0.45]{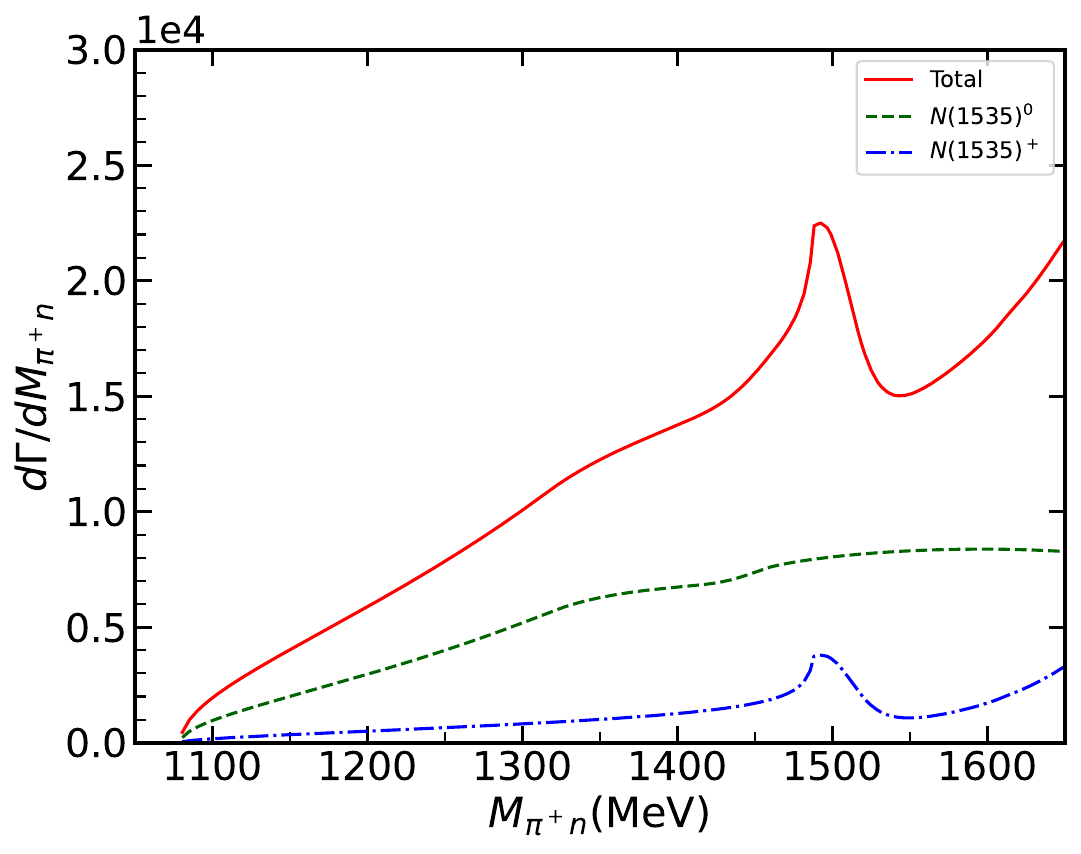}
	\caption{The $\pi^+ n$ invariant mass distribution of the process $\Lambda_c^+ \to \pi^+\pi^0 n$ decay.}
	\label{min13}
\end{figure}
\begin{figure}[tbhp]
	\centering
	\includegraphics[scale=0.45]{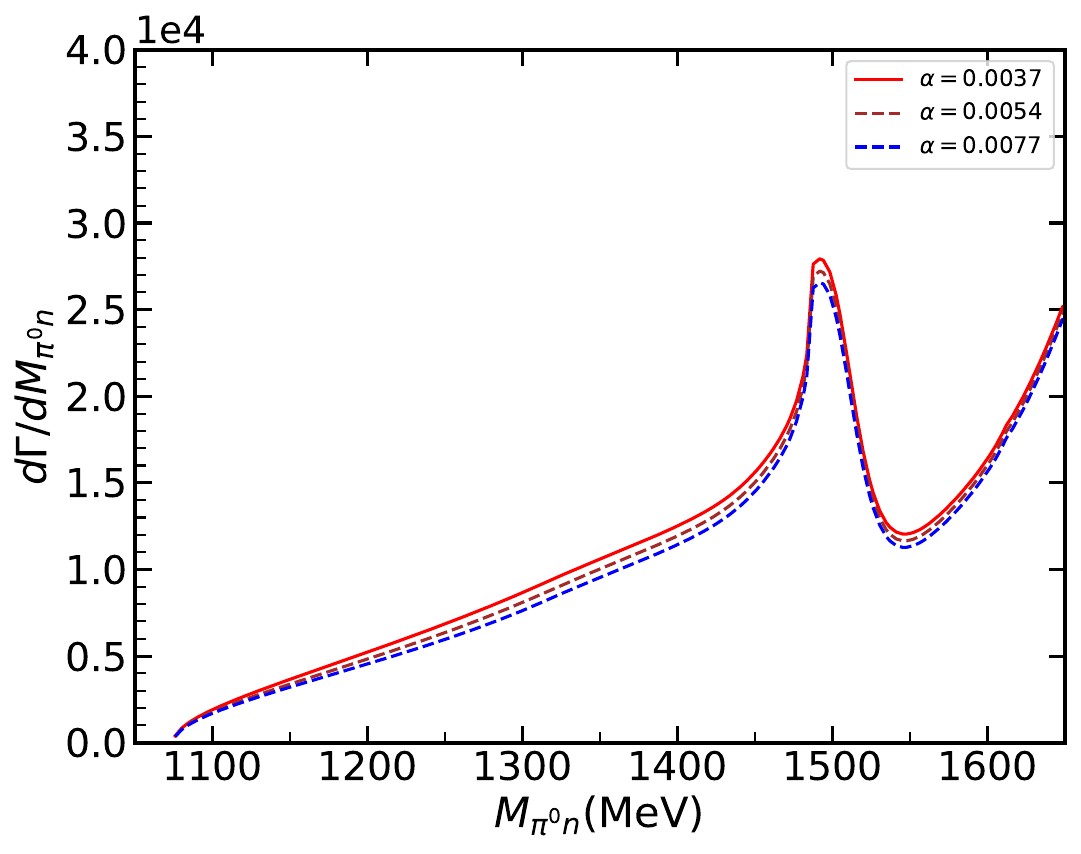}\\
	\includegraphics[scale=0.45]{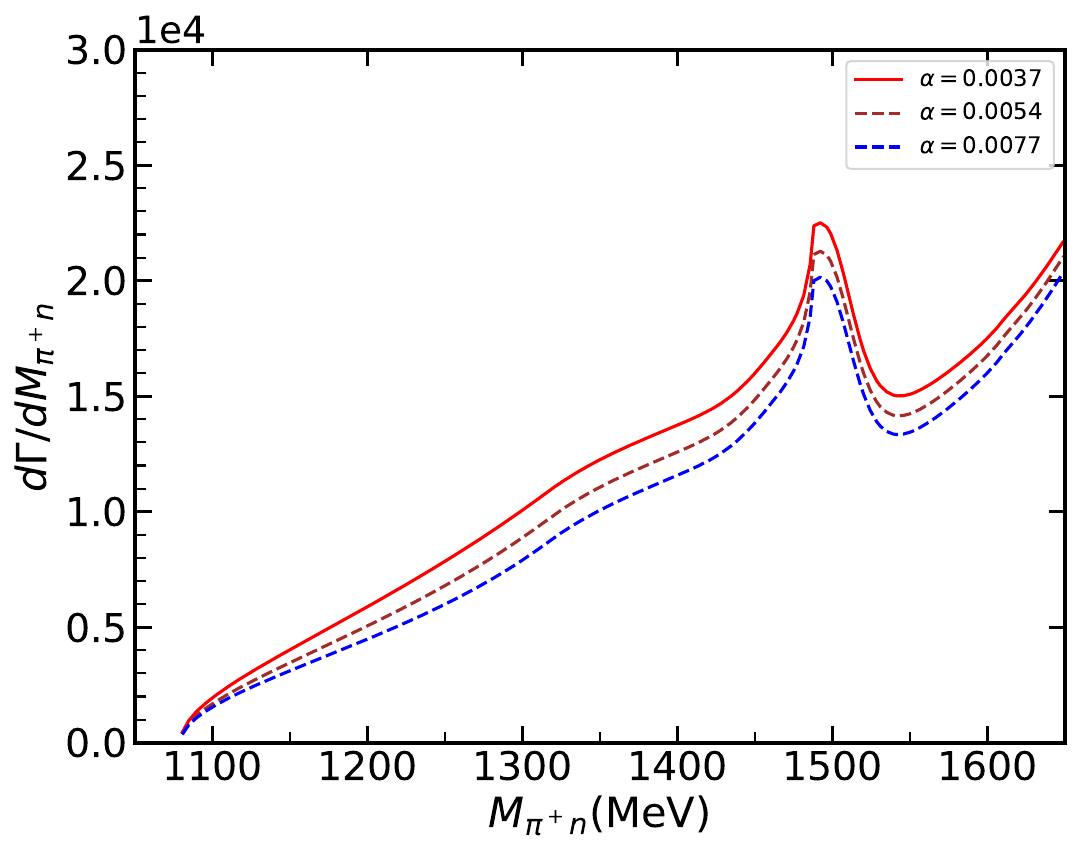}
	\caption{The $M_{\pi^0 n}$ and $M_{\pi^+n}$ distributions with differ-\\ent $\alpha$ values in the smooth factor. The red line, brown dashed and blue dashed curves are obtained with $\alpha =$ $0.0037, 0.0054$ and $0.0077\ \text{MeV}^{-1}$, respectively.}
	\label{fig7}
\end{figure}
\begin{figure}[tbhp]
	\centering
	\includegraphics[scale=0.45]{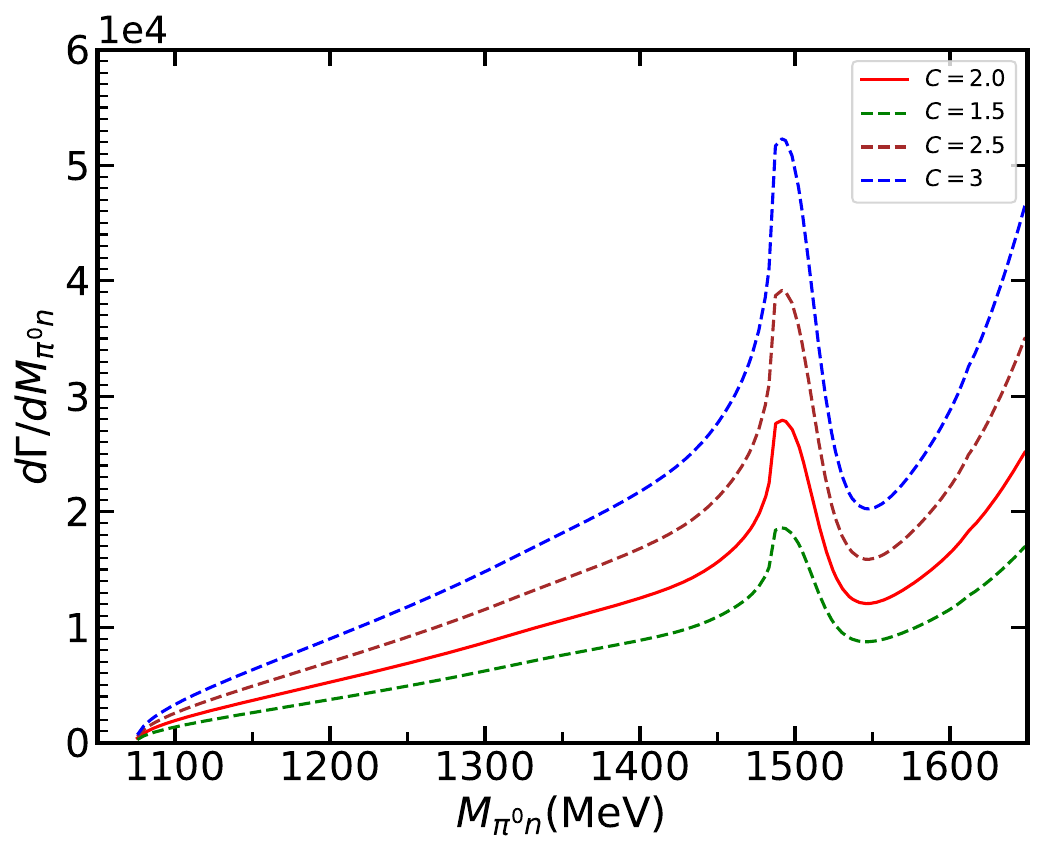}\\
	\includegraphics[scale=0.45]{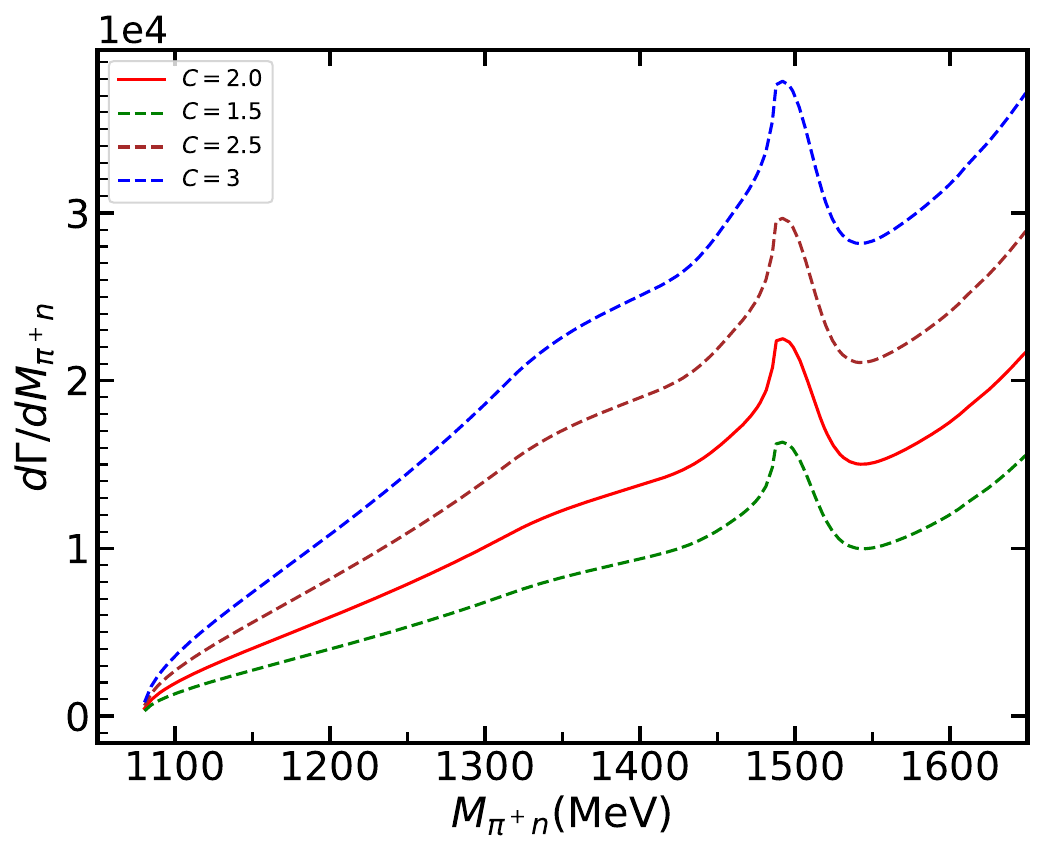}
	\caption{The $M_{\pi^0 n}$ and $M_{\pi^+n}$ distributions for the process $\Lambda_c^+\to \pi^+\pi^0 n$ for different values of color factor $C$.}
	\label{fig8}
\end{figure}
In this section, we present our numerical results based on the theoretical framework established in Sec.~\ref{II}. Our model contains three free parameters: $C$, $V_p$ and $\alpha$. We will use the global normalization factor $V_p=1$ to present the results of arbitrary normalization since it does not affect the shape of the final state invariant mass distribution. The color factor $C$, which represents the relative weight of the external $W^+$ emission mechanism compared to internal $W^+$ emission, is taken to be $C=2$, and we will discuss its dependence for the results later.  The $\alpha$ in the smooth factor added in high energy region in \Eq{eq23}, we set to $\alpha = 0.0037\ \text{MeV}^{-1}$, and the dependence of the results on it is also discussed below.

First, as shown in \FIG{min23} and \FIG{min13}, we show the results of the invariant mass distribution for $\pi^0 n$ and $\pi^+ n$ with $M_{\text{cut}}=1650\ \text{MeV}$, respectively. The green dashed line labeled $N(1535)^0$ and the blue dotted line $N(1535)^+$ correspond to the final state interactions for $\pi^0n$ in \FIG{fig3} and $\pi^+n$ in \FIG{fig4}, respectively. The curve labeled “Total” shows the total contribution result of \Eq{eq19}. 
It is clear from \FIG{min23} and \FIG{min13} that there is a significant peak around $1500$ MeV in both the $\pi^0 n$ and $\pi^+n$ invariant mass distributions, which is associated with the $N(1535)^0$ and $N^+(1535)$ resonance states, respectively. The peak around $1500$ MeV, close to the position of the $N(1535)$ pole in the PDG~\cite{ParticleDataGroup:2024cfk}($1510$ MeV). 

Then we tested the dependence of our results on the parameter $\alpha$ in smooth factor of \Eq{eq23}. As in Ref.~\cite{Debastiani:2016ayp}, we have calculated the invariant mass distributions by taking $\alpha =0.0037\ \text{MeV}^{-1}, 0.0054\ \text{MeV}^{-1}$ and $0.0077\ \text{MeV}^{-1}$, and the results are shown in \FIG{fig7}.
From the invariant mass distributions of $M_{\pi^0 n}$ and $M_{\pi^+n}$ in \FIG{fig7}, we can see that the value of $\alpha$ has only minor effect on the results and does not change the position of the peaks.

Furthermore, in \FIG{fig8}, we compare the invariant mass distributions of $M_{\pi^0n}$ and $M_{\pi^+n}$ under different color factor values of $C=1.5, 2, 2.5, 3$. The result indicates that although variations in the $C$ value affect the strength of the $N(1535)^0$ and $N(1535)^+$ signals, the shape of signals are still evident.

\begin{figure}[tbhp]
	\centering
	\includegraphics[scale=0.45]{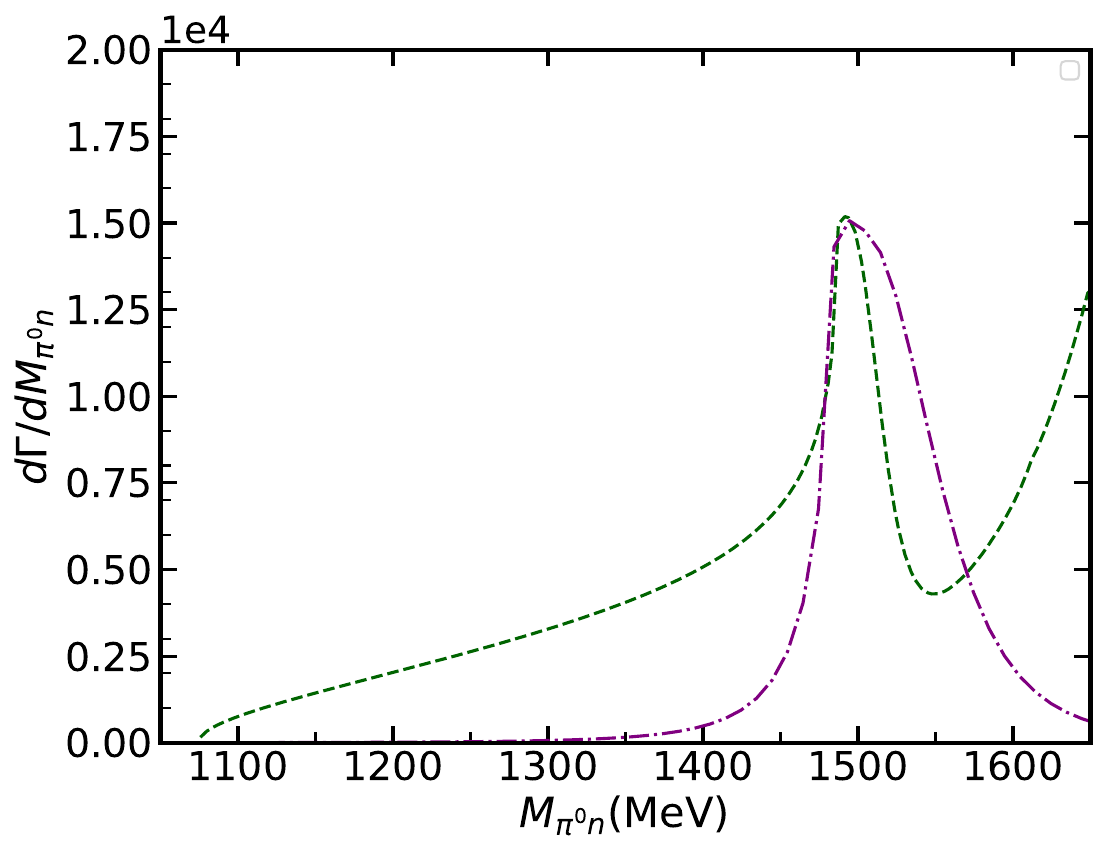}\\
	\caption{The $M_{\pi^0 n}$ distributions for the process $\Lambda_c^+\to \pi^+\pi^0 n$. The green dashed and purple dashdot curves represent the results obtained in Model I, and II, respectively.}
	\label{fig10}
\end{figure}

In \FIG{fig10}, we show the $\pi^0n$ invariant mass distribution, where the green dashed and purple dashdot curves represent the results obtained in Model I and II, respectively. 
Model I is the $N(1535)$  is dynamically generated via the meson baryon interaction as shown in \FIG{fig3}. Model II is the Breit-Wigner amplitude of $N(1535)$ resonance and the coefficients $A$ and $B$ are normalized to the peak of Model I. From the results of Model I and Model II shown in \FIG{fig10}, we see that these two different descriptions of $N(1535)^0$ resonance give different line shapes for the invariant $\pi^0 n$ mass distributions. For the $N(1535)$, the amplitude square obtained with the chiral unitary approach does not behave like an usual Breit-Wigner resonance,
where the resonant shape of Model II is broader than the result of Model I. Future experimental measurements are expected to test our predictions and clarify this issue.

 \section{SUMMARY}
 \label{IV}

The nonleptonic weak decays of charmed baryons serve as a important platform for investigating low-lying excited baryons and also contributes to our understanding of the properties of light baryons with quantum numbers $J^P=1/2^-$. In this work, we examine the $\Lambda_c^+ \to n\pi^+\pi^0$ process within the chiral unitary approach, where the intermediate $N(1535)$ resonance is dynamically generated through $S$-wave interactions between pseudoscalar mesons and baryons by taking into account the $\pi N$, $\eta N$, $K\Lambda$ and $K\Sigma$ channels in the $I=1/2$ sector. 

Based on our calculations, clear peak can be seen around $1500$ MeV in the invariant mass distributions of $\pi^0 n$ and $\pi^+n$, which could be related to the $N(1535)^0$ and $N(1535)^+$ resonances, respectively.
Additionally, due to the limitations of the energy range used in the chiral unitary approach, we progressively suppress the loop functions and amplitudes for $M_{\text{cut}}$ above 1650 MeV. We also discussed the uncertainties in the results by varying the parameter $\alpha$ in the smooth factors. We observe that different $\alpha$ values have a minor effect on the invariant mass distribution, and the peak associated with the $N(1535)$ resonance are still visible. Finally, we also discussed the impact of different color factor $C$ values, which implies that the signal of $N(1535)$ is clearly observable in the invariant mass distribution.
Further experimental measurements are needed to verify the properties of the $N(1535)$ resonance predicted by our model.
 
 \begin{acknowledgments}
 	\noindent
We would like to acknowledge the useful guidance with Prof.\ Eulogio Oset. 	
HS is supported by the National Natural Science Foundation of China (Grant No.12075043). XL is supported
by the National Natural Science Foundation of China under Grant No.12205002.
 \end{acknowledgments}

\bibliography{ref}
\end{document}